# Measurement and Modeling on Terahertz Channel Propagation Through Vegetation

Jiayuan Cui, Yuheng Song, He Jiang, Chenxi Wang, Mingxia Zhang, Guohao Liu, Da Li, Jiabiao Zhao, Jiacheng Liu, Yue Su, Wenbo Liu, Peian Li, Daniel M. Mittleman, *Fellow, IEEE*, Fei Song, *Member, IEEE*, Jianjun Ma, *Member, IEEE*

*Abstract*—The terahertz (THz) band offers promising opportunities for high-capacity wireless communications but faces significant challenges from vegetation-induced channel impairments. This article presents a comprehensive investigation of THz channel propagation through vegetation, introducing a hybrid modeling approach that combines deterministic vegetation-dependent exponential decay (VED) modeling with statistical characterization of temporal variations. Through extensive laboratory measurements using Epipremnum aureum, we find that vegetation introduces angular-dependent power losses, with channel statistics following heavy-tailed Stable distributions rather than conventional Rician or Weibull models. Our outdoor measurements with dense and sparse lilac scenarios reveal pronounced foliage density variations in attenuation and height-dependent effects, while validating the VED model's ability to maintain excellent agreement with measured data and parameter stability across different heights without coefficient recalibration. Critical bit-error-rate (BER) analysis uncovers distinct SNR thresholds beyond which performance exhibits oscillatory behavior due to heavy-tailed fading, establishing fundamental capacity bounds with significant implications for modulation scheme selection and power control strategies in practical THz communication systems.

*Index Terms*—Terahertz channel, Vegetation, Power profile, Bit-error-rate.

## I. INTRODUCTION

THE exponential growth in wireless data traffic, driven by emerging applications including holographic communications, extended reality (XR), and autonomous systems, is pushing wireless networks toward unprecedented capacity demands that are expected to exceed Tbps [1]. While the fifth-generation (5G) networks have successfully leveraged millimeter-wave frequencies, achieving such a high capacity necessitates exploring higher frequency bands, particularly the terahertz (THz) spectrum spanning 0.1-10 THz [2], [3]. The THz band offers several compelling advantages including ultra-wide available bandwidths of tens or even hundreds of GHz, enabling massive data throughput with sub-millisecond latencies critical for real-time applications [4]. These characteristics make THz communications particularly promising for high-capacity wireless backhaul and fixed wireless access networks. However, the practical deployment of THz systems faces significant propagation challenges. The inherently lower diffraction at THz frequencies combined with narrow beamwidth leads to high sensitivity to channel blockage, while the extremely short wavelengths result in enhanced scattering from surface roughness and small obstacles, significantly impacting channel stability and channel reliability [5]. Among the key issues in outdoor scenarios, vegetation-induced channel impairments represent a critical concern, as vegetation is an inevitable and essential element in our daily life [6], [7]. Vegetation can cause substantial attenuation through absorption and scattering, as well as inducing temporal variations in the received signal [8], [9]. Understanding and accurately modeling these multifaceted effects, including both static and dynamic components, is essential for robust system design and reliable deployment in real-world environments.

Channel propagation through vegetation has been extensively characterized at frequencies up to THz band, with studies revealing increasingly complex interactions as wavelengths become comparable to vegetation feature sizes [9]. At millimeter-wave frequencies, vegetation effects exhibit strong dependencies on multiple factors including foliage density, polarization state, and seasonal variations due to changes in moisture content [10]-[12]. Studies have demonstrated that parameters such as leaf thickness, spacing between leaves, and transmission frequency significantly influence both transmitted and reflected fields [13]. Recent advances in electromagnetic parameter extraction techniques have enabled more precise characterization of vegetation's THz responses [14], [15], while revealing that the THz band's high sensitivity to water absorption makes it particularly useful for

This work was supported in part by the National Natural Science Foundation of China under Grant (62471033, 92167204, 62072030), the Science and Technology Innovation Program of Beijing Institute of Technology (2022CX01023), the Fundamental Research Funds for the Central Universities (2024CX06099) and the Talent Support Program of Beijing Institute of Technology "Special Young Scholars" (3050011182153). Corresponding authors: Jianjun Ma, Fei Song.

Jiayuan Cui, Yuheng Song, Mingxia Zhang, Da Li, Jiabiao Zhao, Jiacheng Liu, Wenbo Liu, Peian Li are with Beijing Institute of Technology, Beijing, 100081 China (e-mail: 3120231316, 3220242109, 3120245902, 3220221512, 3120235892, 3120221332, wenbo_liu, 7520240281@bit.edu.cn).

He Jiang and Chenxi Wang is with Beijing Radio Metrology Testing Institute, Beijing, 100039 China (e-mail: jianghe1021@126.com, wangchenxi32@foxmail.com).

Guohao Liu, Yue Su, Jianjun Ma are with Beijing Institute of Technology, Beijing, 100081 China, and Tangshan Research Institute, BIT, Tangshan, Hebei, 063099 China (e-mail: 3220221571; 3220231902; jianjun ma@bit.edu.cn).

Daniel M. Mittleman is with School of Engineering, Brown University, Providence, RI, 02912 USA (e-mail: daniel_mittleman@brown.edu).

Fei Song is with Beijing Jiaotong University, Beijing, 100044 China (e-mail: fsong@bjtu.edu.cn).

monitoring water distribution in plant leaves [16]. This characteristic has proven valuable for evaluating scattering effects and understanding leaf water status [17], with detailed observations of transmission loss and attenuation across different frequency regions providing insights into the effects of leaf thickness and water content variations [18]. These findings underscore the complexity of vegetation-induced channel effects and highlight the need for sophisticated modeling approaches that can capture both macro-scale attenuation and micro-scale scattering phenomena [19].

The evolution of channel models reflects our growing understanding of these complex phenomena. Early models, such as the foundational ITU model, characterized power loss using a simple power-law relationship $L(dB) = Af^B d^C$ [20], where $f$ represents frequency, $d$ denotes vegetation depth, and $A, B, C$ are empirically determined fitting parameters. This framework was subsequently refined through the COST 235 model (valid up to 95 GHz) [21] and the modified exponential decay (MED) model [22], [23] for dense vegetation scenarios. The maximum attenuation (MA) model, later adopted by International Telecommunication Union (ITU), offers an alternative formulation for frequencies up to 30 GHz [23], [24]. However, these models share a fundamental limitation – they cannot adequately account for vegetation (foliage) density variations [25], [26]. Subsequent research revealed that foliage density often exerts greater influence on channel attenuation than specific species characteristics or leaf morphology [27], [28]. This recognition of foliage density's dominant influence led to the formulation of the vegetation-dependent exponential decay (VED) model [29], which explicitly incorporates foliage density ($p$) as a primary propagation parameter

$$L(p,d,f) = Af^B d^C p^D \tag{1}$$

where the exponent $D$ functions as a fourth regression parameter that quantifies the non-linear relationship between foliage density and signal attenuation. This parametric extension represents a significant theoretical advancement over classical models by establishing a direct mathematical coupling between electromagnetic propagation and volumetric scattering properties of the vegetation medium. The VED model demonstrates exceptional promise for THz frequencies, where wavelength-to-foliage size ratios enter a critical regime where neither geometric optics nor homogeneous medium approximations remain valid. Furthermore, this formulation effectively captures averaged power characteristics across heterogeneous vegetation structures, enabling accurate prediction of path loss under varying seasonal and environmental conditions without requiring species-specific calibration.

Despite these modeling advances, critical gaps remain in understanding THz channel propagation through vegetation. The extremely short wavelengths in the THz band ($\lambda < 3$ mm) introduce complex wave-matter interactions that require more sophisticated characterization beyond averaged power levels. Additionally, the ultra-wide bandwidths available at THz frequencies demand comprehensive channel characterization across the entire operational spectrum [19]. To address these challenges, we propose a hybrid modeling approach that combines deterministic and statistical components. The deterministic component captures macro-scale effects such as averaged attenuation, while the statistical component characterizes small-scale fading arising from diffuse scattering [5], [30]. We validate this approach through extensive measurements at 220-230 GHz using two representative vegetation conditions: dense lilac and sparse lilac, chosen for their distinct structural characteristics. Our analysis includes system-level performance evaluation through bit error rate (BER) measurements using standard modulation schemes, providing practical insights for THz communication system design. The remainder of this article is structured as follows: Section II presents our laboratory channel measurements and analysis, including the experimental setup, Markov chain modeling approach, statistical characterization, and VED model validation using Epipremnum aureum vegetation. Section III extends our investigation to outdoor environments, presenting comprehensive measurements with dense and sparse lilac scenarios, seasonal variation analysis, and detailed BER performance evaluation. Finally, Section IV summarizes our findings and discusses future research directions.

## II. CHANNEL PERFORMANCE IN LABORATORY

We developed a high-precision THz measurement system operating from 220-325 GHz to characterize vegetation-induced channel effects (see Fig. 1(a)). The signal generation chain begins with a Ceyear 1465D vector signal generator, producing precise baseband signals from 100 kHz to 20 GHz. These signals undergo frequency up-conversion through a Ceyear 82406D frequency multiplier module, employing an 18× multiplication factor to reach THz frequencies. The radiation system comprises a carefully engineered antenna assembly featuring a Ceyear 89901S horn antenna coupled with a precision-crafted Teflon lens having a focal length of 10 cm. This configuration achieves exceptional directivity with a measured gain of 29 dBi at 230 GHz [31]. The antenna system demonstrates remarkable stability, maintaining a narrow beamwidth of approximately 4° at 230 GHz with minimal pattern deterioration across the measurement bandwidth. Here we set the operating frequency at 230 GHz, since it provides the highest radiation power. At the receiving end, we implemented a symmetrical configuration using a matching horn antenna feeding into a Ceyear 71718 power sensor for direct detection, which has a response time of approximately 50 ms. The power sensor provides high dynamic range measurements with excellent thermal stability, crucial for accurately characterizing vegetation-induced attenuation [32]. System control and data acquisition are managed through a dedicated computer interface operating at a 7 Hz sampling rate, chosen to balance measurement resolution with data management efficiency.

The experimental design implements a bistatic measurement configuration optimized for vegetation characterization, as shown in Fig. 1(a). The transmitter (Tx) and receiver (Rx) are rigidly mounted on an optical bench to ensure positional stability, while the vegetation sample is centrally positioned on a high-precision rotary table between them. Here, we choose the Epipremnum aureus as vegetation sample, for its well-characterized morphological properties and established use in indoor air quality studies [33]. To minimize table reflections and ensure pure vegetation effects, the entire apparatus was positioned 44 cm above the experimental platform, well exceeding the first Fresnel zone (~ 1.59 cm) clearance requirements. This arrangement allows for angular-dependent measurements spanning a complete 360° rotation [34], [35], with measurements taken at 2° increments to provide high

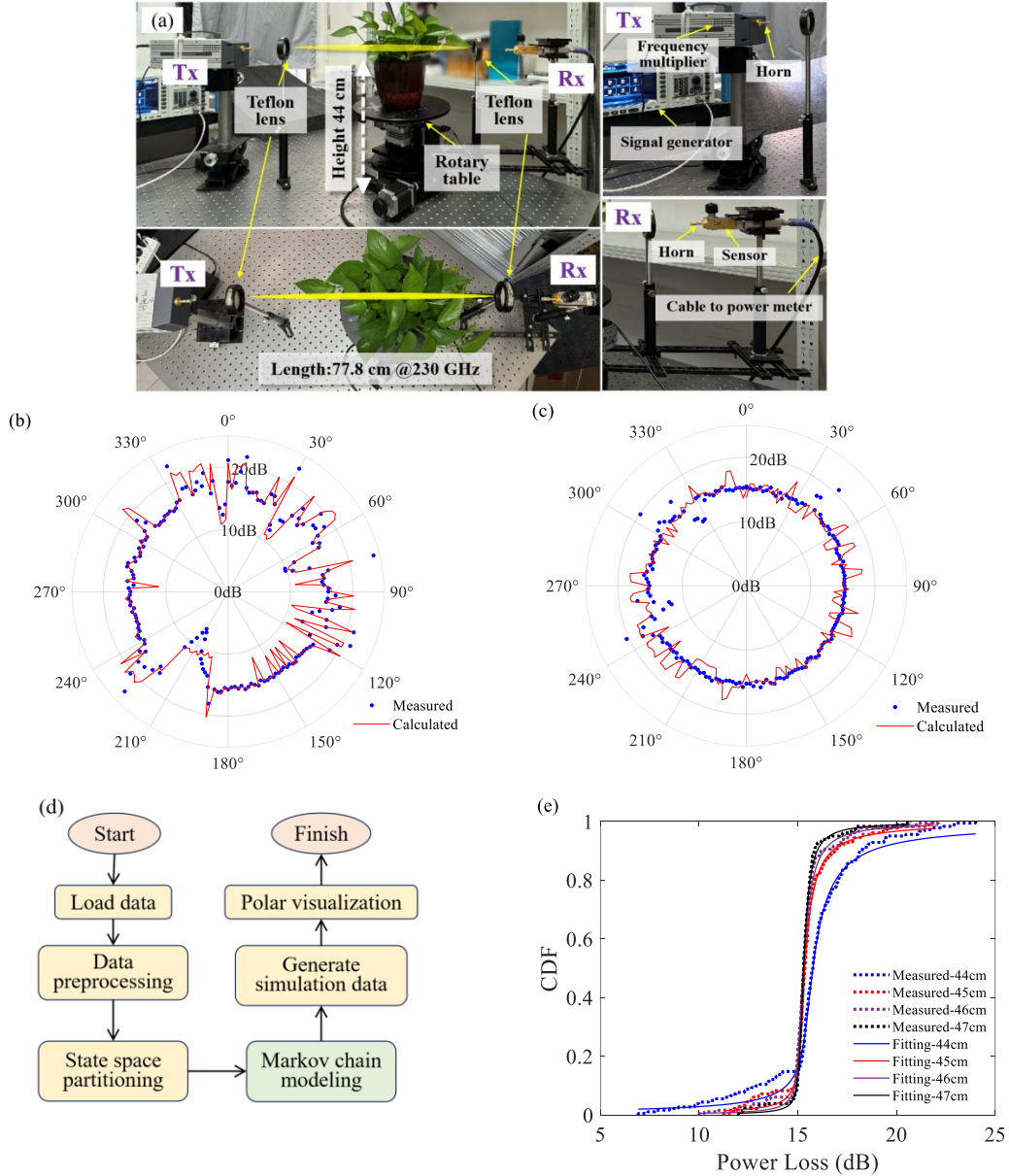

**Fig. 1.** (a) Laboratory channel measurement configuration; Measured and calculated power loss under a channel height of (b) 44 cm and (c) 47 cm by Markov chain model; (d) The Markov chain calculation flow. (e) CDF profile of power loss.

TABLE I
MEAN AND RMS VALUES OF POWER LOSS WITH COMPARISON OF FIT GAP (FG) AND FLUCTUATION ERROR (FE) CALCULATED BASED ON SECTION S1 IN SUPPLEMENTARY INFORMATION PART

| Height $h$ [cm] | Mean [dB] | RMS [dB] | Fit Gap | | | Fluctuation Error | | |
|---|---|---|---|---|---|---|---|---|
| | | | Stable | Rician | Weibull | Stable | Rician | Weibull |
| 44 | 15.792 | 15.994 | 0.0041 | 0.0299 | 0.0379 | 0.0133 | 0.0511 | 0.0573 |
| 45 | 15.512 | 15.584 | 0.0021 | 0.0462 | 0.0895 | 0.0118 | 0.0639 | 0.0786 |
| 46 | 15.405 | 15.464 | 0.0014 | 0.0453 | 0.091 | 0.0102 | 0.0678 | 0.0869 |
| 47 | 15.388 | 15.420 | 0.0009 | 0.051 | 0.1231 | 0.0066 | 0.0691 | 0.0965 |

angular resolution characterization of the vegetation effects.

Our experimental protocol began with rigorous reference measurements in a vegetation-free (free space) environment to establish system performance metrics and reference power levels. These measurements served dual purposes: confirming system linearity and providing essential calibration references for subsequent vegetation measurements. Following reference characterization, we conducted a systematic series of measurements with the vegetation samples fixed on the rotary table. The channel height was varied from $h = 44$ cm to $h = 47$ cm in precise 1 cm increments. This specific height range was chosen to intersect the most dense portions of the vegetation while maintaining consistent beam geometry. At each height setting, we performed complete 360° rotational scans, with the vegetation specimen mounted on the precision rotary table.

*A. Channel measurement*

For each measurement configuration, we determined both mean and root mean square (RMS) values of the power loss. We first detected the received power in the vegetation-free environment to establish a reference, then placed vegetation in the specified position for rotation measurements. The vegetation-induced power loss was obtained based on the difference between these measurements, as illustrated in Fig. 1(b) and (c). Analysis revealed that power loss at most angular positions remains below 20 dB, exhibiting marked directional dependencies that reflect the inherent structural anisotropy of the vegetation. Notably, the power loss shows significant angular dependence due to variations in foliage density at the same height, underscoring that foliage density is a crucial parameter for developing realistic vegetation channel models [27]. This angular dependency and varying power loss patterns suggest that the received power can be effectively described as a random process. Based on this, we hypothesized that the THz channel behavior could be modeled using a Markov chain approach, treating the channel as a discrete-state random process where transitions between power levels depend only on the current state [36], [37]. The state boundaries for this model can be carefully determined through detailed percentile analysis of the measured data, ensuring balanced state occupancy while maintaining physical significance in the power level transitions.

*B. Markov chain modelling*

For the Markov chain implementation, we developed a finite-state stochastic process model with five discrete states to characterize the temporal variations in channel power levels. The computational framework, illustrated in Fig. 1(d), employs a non-uniform state quantization methodology where state boundaries were determined through percentile analysis of the empirical power distribution function. This technique ensures optimal state occupancy equilibrium while preserving the statistical significance of transition events. The state transition probability matrix was constructed by systematically analyzing temporal correlations between consecutive power measurements, thereby capturing both the first-order memory and ergodic properties intrinsic to vegetation-induced channel dynamics. For model parametrization, we utilized the measurement data at $h = 44$ cm to derive the following transition probability matrix

$$P = \begin{bmatrix} p_{11} & p_{12} & p_{13} & p_{14} & p_{15} \\ p_{21} & p_{22} & p_{23} & p_{24} & p_{25} \\ p_{31} & p_{32} & p_{33} & p_{34} & p_{35} \\ p_{41} & p_{42} & p_{43} & p_{44} & p_{45} \\ p_{51} & p_{52} & p_{53} & p_{54} & p_{55} \end{bmatrix}$$
$$= \begin{bmatrix} 0.4167 & 0.1667 & 0.1667 & 0.1111 & 0.1389 \\ 0.0857 & 0.4286 & 0.1717 & 0.2571 & 0.0571 \\ 0.1892 & 0.1892 & 0.2432 & 0.2703 & 0.1081 \\ 0.1667 & 0.1667 & 0.2778 & 0.2778 & 0.1111 \\ 0.1111 & 0.0556 & 0.1667 & 0.0833 & 0.5833 \end{bmatrix} \quad (2)$$

where each element $p_{ij}$ quantifies the conditional probability of transitioning from state $i$ to state $j$ in the subsequent sampling interval. This matrix encapsulates the complete Markovian behavior of the channel.

Each state in the Markov model was characterized by a normal distribution, with the mean value centered on the corresponding power level. The standard deviation was optimized to 0.05 times the state width to accurately represent the observed variations. This hybrid approach, combining discrete state transitions with continuous intra-state distributions, offers a more nuanced representation of the channel dynamics compared to traditional binary or purely discrete models.

Validation of the Markov chain model was performed using measurements at a channel height of $h = 44$ cm. As shown in Fig. 1(b), the model demonstrates excellent agreement with measurement data across all angles, supporting its accuracy in characterizing the channel behavior at this specific height. However, when applied to measurements at a different height of $h = 47$ cm (Fig. 1(c), significant discrepancies emerge, indicating that the Markov chain model's validity does not extend across different channel heights. This limitation was previously observed in V-band channel measurements and can be attributed to variations in foliage density [38]. This conclusion is further supported by the systematic changes in mean and RMS values of power loss observed at different channel heights, as detailed in Table I, where both metrics show a decreasing trend with increasing channel height.

*C. Statistical analysis*

To gain deeper insights into the channel characteristics, we analyzed the cumulative distribution function (CDF) of the power loss, as shown in Fig. 1(e). The channel exhibits distinctive heavy-tailed statistical characteristics, resulting from complex interactions of multipath scattering components. These heavy tails indicate that the channel experiences occasional severe fades or interference bursts, which can significantly impact channel performance [39].

As evaluated in Table I, the Stable distribution model proves superior in describing these fading characteristics and statistical anomalies compared to alternative distributions. This model effectively captures asymmetric fading patterns (where signal variations are nor evenly distributed around the mean value, unlike in normal distributions, due to non-uniform foliage density) and accurately represents extreme fading events [40]. It should be noted that, despite the vegetation's overall obstructive effect on channel transmission, our analysis reveals opportunities for transmission optimization. The observed variations in power loss at different angles indicate non-uniform foliage density, suggesting that certain propagation paths may offer superior transmission characteristics. This finding has practical implications for THz communication system deployment - by carefully analyzing these angular dependencies and adjusting equipment positioning to exploit areas of lower foliage density, channel performance can be significantly enhanced. This observation also highlights a key limitation of the Markov chain model, which cannot capture these spatial optimization opportunities, indicating the need for more sophisticated modeling approaches that can account for spatial foliage density variations.

*D. VDE channel modelling*

Given the limitations of the Markov chain model in handling height-dependent channel variations, we investigated the VED model (Eq. (1)) to better characterize the THz channel through vegetation. This model explicitly incorporates foliage density through the plant area index (*PAI*) parameter $p$, which provides a quantitative measure of the vegetation's structural characteristics, as

$$PAI = LAI + WAI \qquad (3)$$

where *LAI* represents the one-sided leaf area per unit ground surface area and *WAI* denotes the woody component contribution (branches and stems) per equivalent area [41]. This biophysical parameterization effectively captures the vegetation's volumetric scattering and absorption properties at THz frequencies.

For precise *PAI* determination, we employed a Gap Light Analyzer with hemispherical photography following the standardized optical transmission methodology detailed in [42], as illustrated in Fig. 2(a). This non-destructive technique ensures reproducible characterization of foliage density through analysis of canopy gap fraction distributions. Parameter estimation was conducted through constrained nonlinear optimization using the *nonlinear regression* algorithm of MATLAB@2024a, applied to our reference measurement dataset ($h$ = 44 cm, vegetation depth $d$ = 25 cm). The optimization yielded coefficients $A$ = 9.3148, $B$ = 0.1239, $C$ = 0.8077, and $D$ = 0.7639, with complete parametric details provided in Table SI of the Supplementary Information.

As demonstrated in Fig. 2(b), the model exhibits exceptional agreement with measured power loss profiles at the reference height. Most significantly, when applied to measurements at alternative heights ($h$ = 45, 46, and 47 cm) with consistent vegetation depth ($d$ = 25 cm) but varying foliage density distributions (detailed in Section S3 of Supplementary Information and visualized in Fig. 2(c)), the model preserved its predictive accuracy without coefficient recalibration. This height-invariant performance validates the model's ability to capture the fundamental physics of THz-vegetation interactions through the *PAI* parameter, demonstrating superior robustness compared to stochastic approaches when characterizing complex heterogeneous propagation environments.

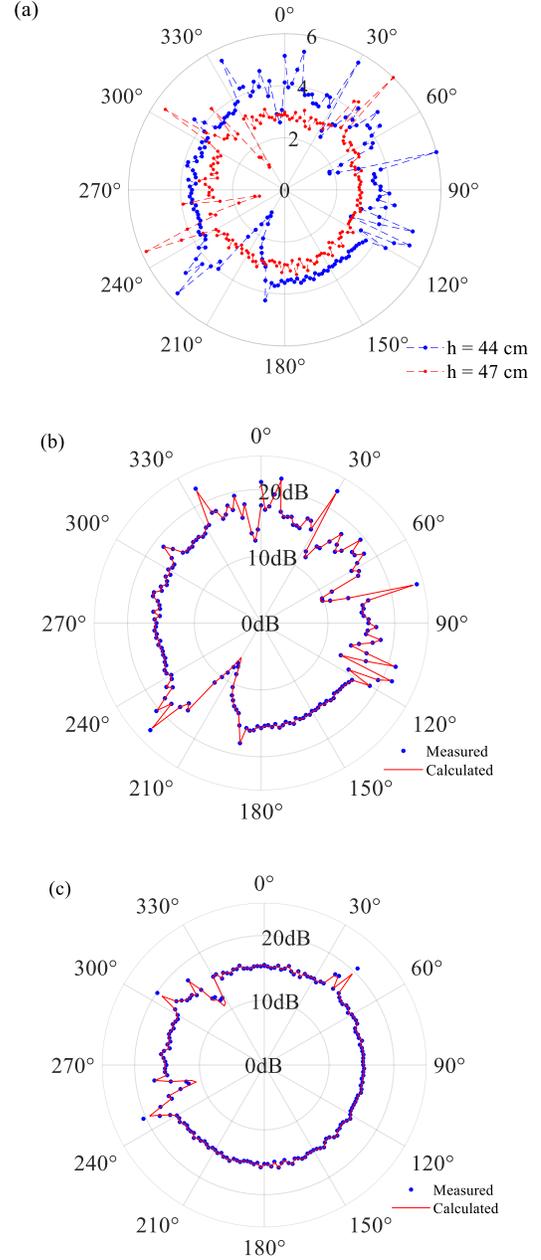

**Fig. 2.** (a) Measured *PAI* values using a Gap Light Analyzer. Comparison between measured and calculated (using VED model) power losses under a channel height of (b) 44 cm and (c) 47 cm.

This height independence represents a significant advancement over the Markov chain approach, as it suggests that the VED model captures fundamental physical relationships between vegetation characteristics and channel behavior. The model's success can be attributed to its incorporation of *PAI*, which effectively quantifies the vegetation's spatial distribution and density [43]. This finding aligns with previous studies showing that foliage density often exerts greater influence on channel attenuation than specific morphological features [27], [44]. Based on these promising results, we adopted the VED model as our primary analytical framework for subsequent channel characterization across different vegetation scenarios.

To validate the VED model's applicability across different measurement systems, we conducted additional verification using a high-precision data link operating at 102 GHz, as illustrated in Fig. 3(a). This heterodyne transceiver architecture implemented a dual-conversion topology to enable modulated signal characterization beyond simple power measurements. The transmitter subsystem employed a Tektronix AWG70002A arbitrary waveform generator (sampling rate: 25 GSa/s, DAC resolution: 10 bits) to synthesize a 2 GHz intermediate frequency (IF) signal with quadrature phase-shift keying (QPSK) modulation at 1 GSa/s symbol rate. A root-raised cosine filter with roll-off factor of 0.35 provided spectral shaping, yielding an effective signal bandwidth of 1.35 GHz. This baseband signal was upconverted using a Sinolink SLFS440D microwave synthesizer generating a 34 GHz carrier, which underwent frequency triplication to produce the local oscillator for the final mixing stage, resulting in a 102 GHz RF signal. The RF chain incorporated a high-gain power amplifier to compensate for free-space propagation losses, while ferrite isolators suppressed spurious reflections and maintained impedance matching throughout the transmission path. A precision horn antenna with optimized directivity completed the transmitter front-end. At the receiver terminal, an identical horn antenna captured the spatially propagated signal before down conversion through a reciprocal mixing process. The 102 GHz RF signal was heterodyned to the 2 GHz IF band, filtered through a precision low-pass network, and acquired using a Tektronix 75902SX digital oscilloscope (bandwidth: 59 GHz, sampling rate: 200 GSa/s). Vector signal analysis was performed using the integrated SignalVu VSA software package, enabling comprehensive demodulation and constellation analysis.

For vegetation measurements, we established an 80 cm propagation channel with Epipremnum aureum vegetation (25 cm height) positioned at the midpoint. Due to receiver mechanical constraints, angular characterization was limited to discrete rotation positions of the vegetation sample, as depicted in Fig. 3(b). Comparative analysis between measured power loss data and VED model predictions (using angular-dependent *PAI* values documented in Table SII of the Supplementary Information) demonstrated excellent correlation, further validating the model's accuracy across different measurement methodologies. It should be noted that this vegetation specimen differs from that utilized in the experiments described in Fig. 1, as measurements were conducted approximately one year apart, during which the Epipremnum aureum exhibited significant morphological development. This modulated signal measurement system served dual analytical purposes: beyond power loss characterization, it enabled comprehensive bit error evaluation, which will be examined in detail in the subsequent section to establish system-level performance metrics under vegetation-impaired channel conditions.

## III. CHANNEL PERFORMANCE IN OUTDOOR

### A. Channel measurement

To further investigate vegetation effects on THz channel propagation, we conducted comprehensive outdoor channel measurements at the Liang Xiang Campus of Beijing Institute of Technology using lilac (Syringa vulgaris) as the representative vegetation specimen. The choice of lilac was based on its well-defined structural characteristics with an average height of 1.8 m and leaf dimensions comparable to the wavelength at THz frequencies, making it suitable for characterizing vegetation-induced channel effects. Based on the experimental setup as in Fig. 1(a), we further employed a TCM-PA220 power amplifier (~ 24 dB amplification) before the antenna and replacing the teflon lens with the HDPE lens (focal length 30 cm) to increase the output power from the transmitter. This assembly ensures beam gains of 60.5 dBi at 220 GHz, combining the gains from both transmitting and receiving antennas and their respective HDPE lenses. The measurement setup is shown in Fig. 4(a) and the measured signal-to-noise ratio (SNR) for the line-of-sight (LOS) channel (free space) is around 20.3 dB, which means we would have enough SNR for the further channel measurement in vegetation scenarios.

We performed 220 GHz to 230 GHz sweep measurements in dense lilac (Summer) and sparse lilac (Spring) scenarios, with a fixed channel distance of 1.8 m between the HDPE lenses, as the LOS. This frequency is determined by the operating frequency range of the TCM-PA220 power amplifier. For reference measurement at the same distance, we carefully moved the upper portions of the lilac vegetation away from the direct channel path while keeping their root positions fixed, ensuring an unobstructed LOS channel. Two channel heights are selected for each vegetation scenario to analyze the influence of foliage density on channel performance. For the scenario with dense lilac, the height changes from $H_{d1}$ (= 1.25 m) to $H_{d2}$ (= 1.22 m), corresponding to vegetation depths of $d$ =1.8 m and 1.75 m, respectively. For the sparse lilac scenario, the channel height changes from $H_{s1}$ (= 1.25 m) to $H_{s2}$ (= 1.22 m), with corresponding vegetation depths of $d$ =1.36 m and 1.3 m, respectively.

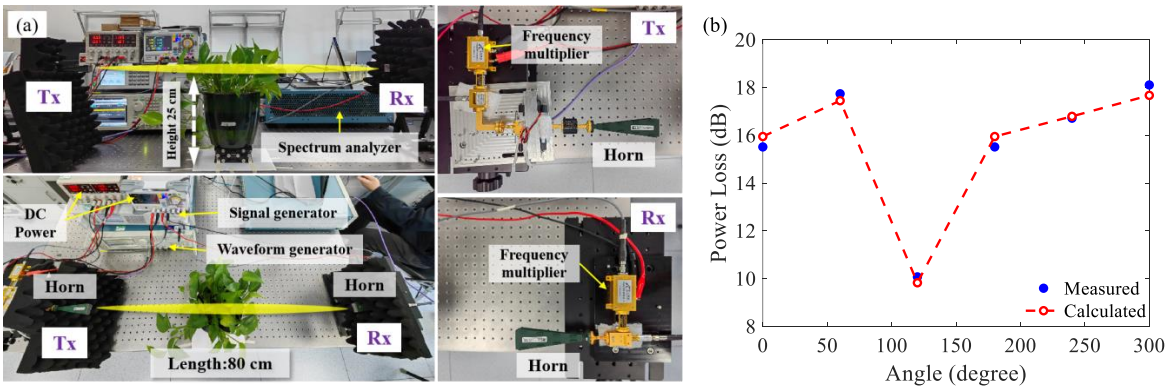

**Fig. 3**. (a) Laboratory channel measurement configuration; (b) Change of power loss with rotation angle.

The measurement result is shown in Fig. 4(b) and (c). It can be seen that there is a large difference under different channel heights and different seasons, due to the variation of foliage density and depth. The difference can be 2.7 dB for height $H_{i2}$ ($i$ =$d$ or $s$) and 4 dB for $H_{i1}$. This means foliage density's variations can result in pronounced channel attenuation. Thus, in practical deployment scenarios, communication systems need to account for foliage density fluctuations, and need adaptive power control mechanisms in system design. Besides, can mitigate some of the signal losses caused by vegetation. This would be particularly important for applications such as fixed wireless access, where consistent performance is needed.

*B. Channel modeling*

For the theoretical calculation, we employ the VED model again by setting the values of coefficients ($A$ = 0.9556, $B$ = 0.1477, $C$ = 2.9908 and $D$ = 0.9703) based on the scenario with the channel height of $H_{d1}$ and setting the values of coefficients ($A$ = 0.9653, $B$ = 0.2107, $C$ = 3.1236 and $D$ = -1.0463) based on the scenario with the channel height of $H_{s1}$ as shown in the Section S5 in the Supplementary Information part. This negative $D$ value, obtained through fitting the VED model to our measured data, indicates reduced power loss with increasing foliage density in some special situations. This counter-intuitive condition may arise from constructive interference effects, when vegetation spacing, wavelength-to-vegetation size ratios, and path geometries align [45]. It may also result from advantageous wave guiding, or beneficial diffraction patterns in sparse vegetation structures [46]. However, our current measurement capabilities do not allow us to definitively identify the underlying physical mechanisms.

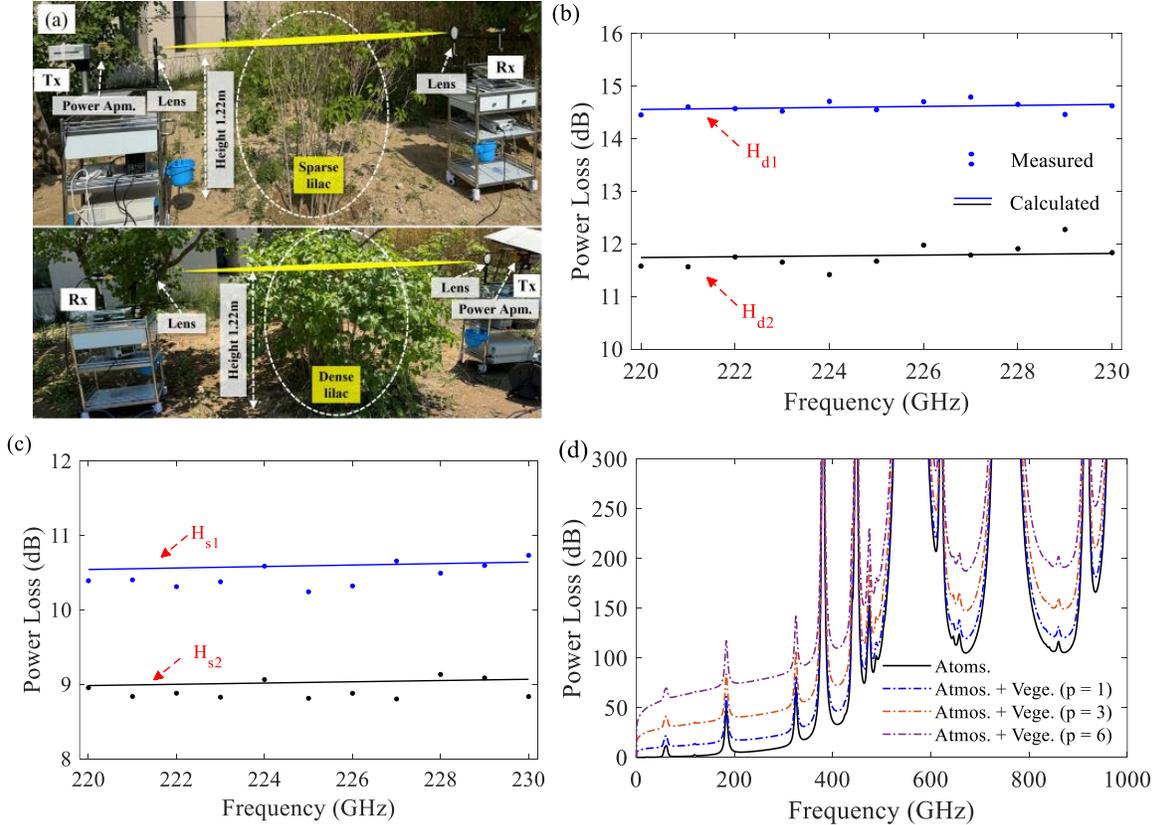

**Fig. 4.** (a) Outdoor channel measurement configuration; Measured and calculated channel power loss in (b) dense lilac and (c) sparse lilac conditions; (d) Calculated spectral power loss due to vegetation ($p$ = 1, 3, and 6) and/or gaseous absorption, using the ITU-R model [48] with temperature 25ºC and humidity RH 60%.

the impact of channel height on channel attenuation suggests that careful planning of antenna placement is crucial. Ensuring that communication channels are deployed at optimal heights

The calculation result is shown in Fig. 4(b) and (c) employing the measured value of *PAI* (using the Gap Light Analyzer) for each specific vegetation scenario. It can be seen that the calculated result agrees well with the measured data, with the discrepancy below 0.7 dB always. This validates the accuracy and reliability of the VED model. This level of precision implies that the model can confidently be used to predict average channel behavior across different heights without re-calibrating the parameters, as long as the vegetation scenario remains consistent. In other words, the stability of the coefficients (*A*, *B*, *C*, and *D*) across different channel heights (Table SIII in Supplementary Information) signifies that vegetation effects can be characterized with a fixed set of parameters derived from a single height measurement. Once these parameters are fitted using a known *PAI* for a given foliage density and type, they remain applicable for height variations. This simplifies the modeling process significantly, as there is no need for repeated empirical measurements or model adjustments when optimizing antenna placement at different heights. For communication system designers, this provides a robust tool to anticipate signal loss due to vegetation across various link configurations.

To assess broadband performance, we extended the model analysis across the 0.1-1 THz spectrum using the dense lilac coefficients ($H_{d1}$ configuration with *PAI* $p$ = 1, 3 and 6), over a channel distance of 1 km. The calculated results, shown in Fig. 4(d), reveal that vegetation (with $p$ = 1) introduces power losses ranging from 3-14 dB across the entire band, besides the atmospheric absorption loss. And when the *PAI* increases to $p$ =6, the power loss increases to the range of 30-86 dB. This observation has important implications for system design, particularly when operating in atmospheric transparency windows where longer propagation distances are feasible. In these windows, vegetation losses become an important factor

Conversely, when operating near atmospheric absorption peaks, the relative impact of vegetation becomes less significant compared to the dominant atmospheric absorption, potentially simplifying system design for specific applications where these frequencies are preferred [47].

*C. Statistical analysis*

To further analyze the channel fading characteristics, we examined the CDF of the SNR for the 220 GHz channel under different vegetation scenarios, as shown in Fig. 5, using the same experimental setup as in Fig. 4(a). Statistical analysis reveals that both dense and sparse lilac environments exhibit fading behavior best characterized by the Stable distribution, as evidenced by the fitting parameters in Table II. The superior fit of the Stable distribution can be attributed to the unique wave-matter interactions at THz frequencies, where wavelengths are comparable to or smaller than typical leaf features. In dense lilac scenarios, multiple scattering from leaves, branches, and stems creates a complex multipath environment, leading to a heavy-tailed distribution of SNR values that deviates from classical Rician or Weibull models. Even in sparse lilac conditions, the spatial heterogeneity of foliage density and partial beam occlusions generate significant SNR variations that are better captured by the Stable distribution's ability to model extreme events and non-Gaussian behavior. This is quantitatively demonstrated by the lower Fitting Gap (FG) and Fluctuation Error (FE) metrics for the Stable distribution across all scenarios (Table II), compared to Rician and Weibull models. This indicates that the channel is less predictable instantly and subject to more extreme performance fluctuations, which can pose challenges for maintaining reliable communication links [49].

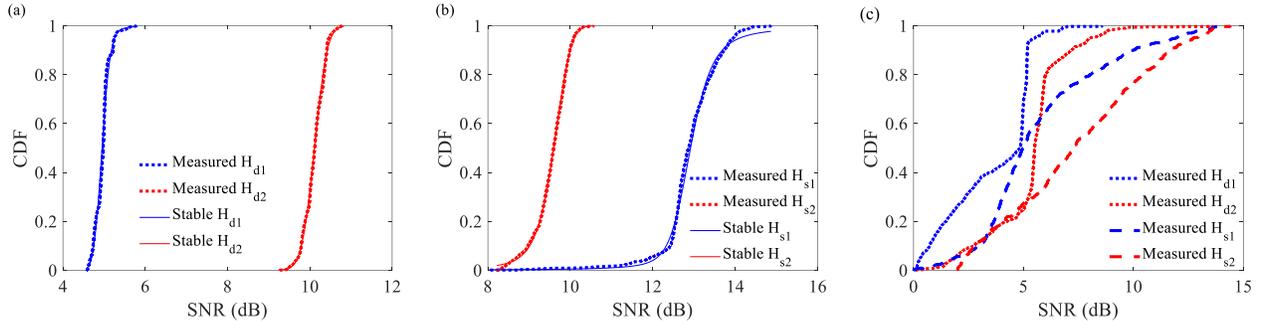

**Fig. 5.** CDF distribution in (a) dense lilac and (b) sparse lilac conditions without wind; (c) CDF distribution in both conditions with wind. (b) keeps the same legend with (a).

TABLE II
COMPARISON OF FITTING GAP AND FLUCTUATION ERROR (FE) FOR DIFFERENT DISTRIBUTIONS

| Scenario | Fitting Gap | | | Fluctuation Error | | |
|---|---|---|---|---|---|---|
| | Stable | Rician | Weibull | Stable | Rician | Weibull |
| Dense lilac ($H_{d1}$) | 0.0137 | 0.0149 | 0.0456 | 0.0235 | 0.0259 | 0.0472 |
| Dense lilac ($H_{d2}$) | 0.0019 | 0.0021 | 0.0054 | 0.0102 | 0.0104 | 0.0157 |
| Sparse lilac ($H_{s1}$) | 0.0052 | 0.0263 | 0.0224 | 0.0119 | 0.0295 | 0.0329 |
| Sparse lilac ($H_{s2}$) | 0.0005 | 0.0115 | 0.0005 | 0.0054 | 0.0189 | 0.0055 |

that must be carefully considered in link budget calculations.

The introduction of wind-induced motion fundamentally alters the channel statistics, as evidenced by the empirical CDF shown in Fig. 5(c). Under wind conditions (velocity magnitude ~10 km/h in both experimental scenarios), the SNR exhibits a complex multi-modal distribution that resists characterization through classical single-parameter Stable distributions. While our measurement system's temporal resolution (7 Hz sampling frequency) imposes bandwidth limitations on capturing high-frequency vegetation-induced variations, this constraint provides the advantageous effect of acting as an implicit low-pass filter, facilitating the delineation between distinct temporal scale regimes in the vegetation's mechanical response. Consequently, our measurements predominantly capture intermediate-to-low frequency components associated with primary branch oscillations and bulk canopy displacement, rather than the higher-frequency micro-scale perturbations from individual leaf flutter and torsional vibrations. This observation underscores the necessity for hierarchical multi-scale statistical frameworks capable of modeling both the quasi-periodic slower variations from structural vegetation movement and the rapid fluctuations arising from dynamic scattering/absorption phenomena, even when instrumentation constraints preclude complete resolution of the full temporal spectrum of channel dynamics.

### D. BER modelling and analysis

While our previous channel characterization revealed the prevalence of heavy-tailed fading statistics, understanding how these channel characteristics impact digital communication performance requires careful examination of error rates under different modulation schemes. The analysis of BER performance provides crucial insights into the reliability and robustness of THz communication systems operating in vegetated environments. Based on the observed Stable distribution characteristics of the channel, we selected QPSK and 16 quadrature amplitude modulation (16-QAM) for comparative analysis. QPSK offers inherent resilience to amplitude variations through its phase-based encoding, while 16-QAM enables higher spectral efficiency at the cost of increased sensitivity to amplitude distortions [50].

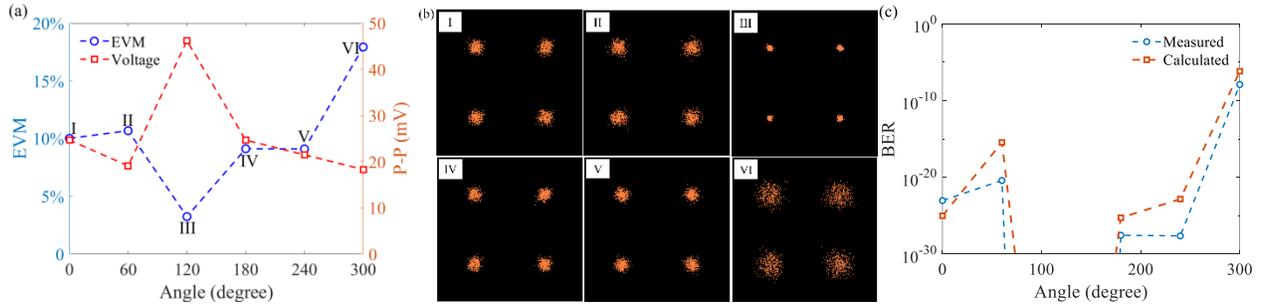

**Fig. 6.** (a) EVM and voltage peak-to-peak values at each rotation angle; (b) Constellation diagrams at each rotation angle; (c) Comparison of BER between measured and calculated values.

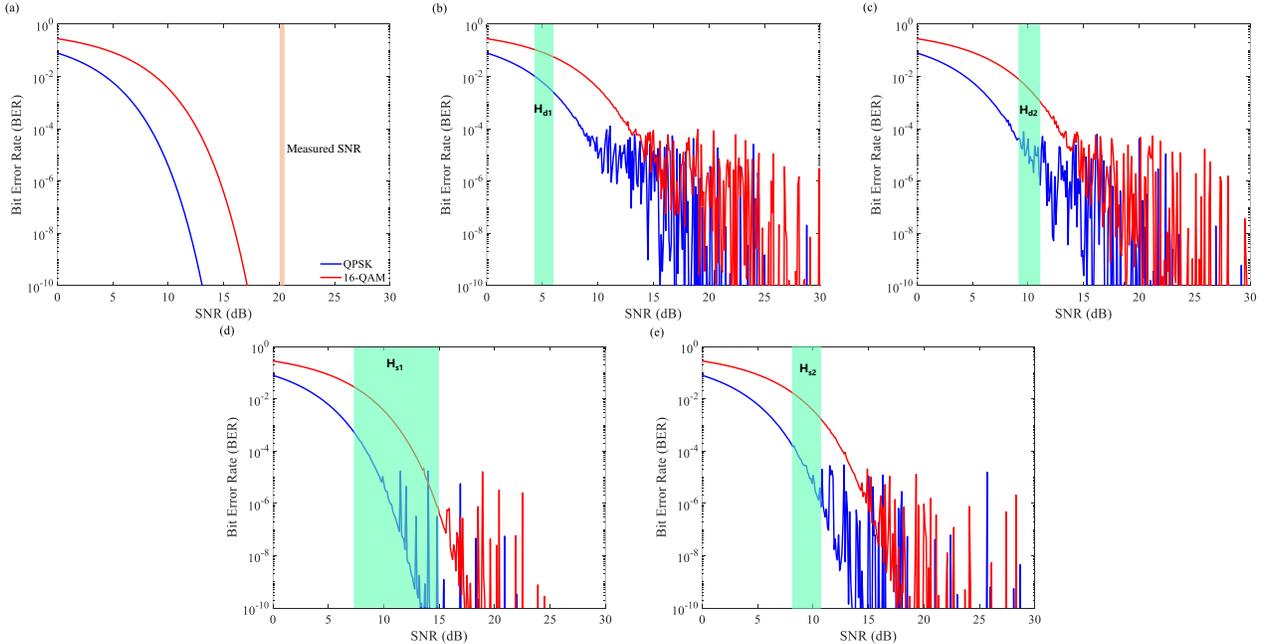

**Fig. 7.** BER performance through (a) free space; dense lilac with a channel height of $H_{d1}$; (c) dense lilac environment with a channel height of $H_{d2}$; (d) sparse lilac environment with a channel height of $H_{s1}$; (e) sparse lilac with a channel height of $H_{s2}$. (f) BER performance with respect to frequency. (b)-(e) keep the same legend with (a).

In the reference case of an unobstructed LOS channel, the system performance can be analyzed using classical additive white Gaussian noise (AWGN) models. For QPSK modulation, the theoretical BER is given by

$$BER_{QPSK}(\gamma) = \frac{1}{2}\left(1 - \sqrt{\frac{\gamma}{1+\gamma}}\right) \quad (4)$$

where $\gamma$ represents the SNR. Similarly, for 16-QAM, the BER follows the formula

$$BER_{16-QAM}(\gamma) = \frac{3}{8}\left(1 - \sqrt{\frac{\gamma}{5+\gamma}}\right) \quad (5)$$

However, in vegetated environments where the channel exhibits heavy-tailed Stable distribution fading, the analysis becomes more complex as the SNR no longer follows a simple probability density function. Instead, we must characterize the BER through the characteristic function of the Stable distribution. For QPSK under Stable fading, the BER becomes

$$BER_{QPSK} = \int_0^\infty \frac{1}{2}\left(1 - \sqrt{\frac{\gamma}{1+\gamma}}\right) \cdot f(\gamma) d\gamma \quad (6)$$

where $f(\gamma)$ represents the probability density function of the instantaneous SNR. This function can be obtained through numerical inversion of the characteristic function

$$f(\gamma) = \frac{1}{2\pi}\int_{-\infty}^{\infty} e^{-it\gamma + it\mu - |\sigma t|^\alpha [1 - i\beta\, sign(t)\Phi(\alpha,t)]} dt \quad (7)$$

In this formulation, $\alpha$ denotes the characteristic exponent governing tail behavior ($0 < \alpha \leq 2$), $\beta$ represents the skewness parameter ($-1 \leq \beta \leq 1$), $\sigma$ quantifies the scale parameter, and $\mu$ corresponds to the location parameter. These parameters, obtained through maximum likelihood estimation and documented in Table SIV of the Supplementary Information, fully characterize the stochastic channel behavior. The stability correction function $\Phi(\alpha,t)$ implements the necessary parameterization adjustment

$$\Phi(\alpha,t) = \begin{cases} \tan(\frac{\pi\alpha}{2}), & \alpha \neq 1 \\ \frac{2}{\pi}\log|t|, & \alpha = 1 \end{cases} \quad (8)$$

For higher-order modulation schemes such as 16-QAM, the error probability extends to

$$BER_{16-QAM} = \int_0^\infty \frac{3}{8}\left(1 - \sqrt{\frac{\gamma}{5+\gamma}}\right) \cdot f_\gamma(\gamma) d\gamma \quad (9)$$

To experimentally validate these theoretical formulations, we conducted comprehensive error vector magnitude (EVM) measurements using the communication testbed depicted in Fig. 3(a), employing QPSK modulation across multiple angular orientations. The results, presented in Fig. 6(a), demonstrate substantial EVM variability corresponding to angular-dependent *PAI* variations, while Fig. 6(b) visualizes the corresponding constellation diagrams at each measurement point. For quantitative comparison with theoretical predictions, we converted measured EVM values to equivalent BER using the established relationship:

$$BER = \frac{1}{2} erfc\left(\frac{1}{\sqrt{2}} EVM\right) \quad (10)$$

where erfc(·) denotes the complementary error function. The theoretical BER, calculated using Eqs. (6-8) with the fitted Stable distribution parameters from Table SIV of Supplementary Information, demonstrates excellent correspondence with the experimentally derived values as shown in Fig. 6(c). This validation confirms the model's accuracy across diverse vegetation conditions and establishes a robust analytical framework for predicting communication system performance in THz vegetation channels, forming the foundation for our subsequent investigation of modulation schemes and system optimization strategies.

Further for the lilac scenario, our calculated results reveal several significant phenomena. In the vegetation-free scenario (see Fig. 7(a) using Eq. (4) and (5)), the system achieves BER levels as low as $10^{-10}$ within our measurement SNR range, following conventional AWGN channel behavior. However, as shown in Fig. 7(b-e) (the green highlight represents the measured SNR range in Fig. 5 (a) and (b)), the presence of vegetation fundamentally alters this behavior, in a remarkable and counterintuitive way. Both dense and sparse lilac scenarios exhibit a critical SNR threshold (10 dB for QPSK, 15 dB for 16-QAM) beyond which the BER performance begins to oscillate rather than monotonically improve. This limitation stems from the heavy-tailed nature of the Stable fading distribution, where the channel experiences extreme variations in signal strength that persist regardless of transmission power. Unlike traditional fading where increasing power helps average out random variation, these heavy-tailed statistics mean that unpredictable deep fades can occur even at high SNR levels. It is analogous to trying to see through moving vegetation - simply making the light source brighter doesn't help if the leaves continue to create random patterns of shadows and openings. This behavior aligns with Koch and Lapidoth's analysis [51] showing that for this heavy-tailed fading (Stable fading) channels with sporadic and unpredictable path gain variations, capacity becomes bounded at high SNR. Our calculations using real-world fading parameters provide a persuasive numerical validation of this theoretical prediction. This finding can provide necessary implications for designing reliable THz communication systems in vegetated or similar environments. Rather than pursuing higher transmission powers, we should focus on adaptive techniques like dynamic coding schemes that can respond to channel statistics, spatial diversity approaches that exploit multipaths, and intelligent routing strategies that can identify and utilize paths with lower channel fading.

The consistency of these critical SNR thresholds across different vegetation densities and channel heights suggests a fundamental limitation rather than an environmental effect. This observation further confirms the conclusion that capacity may be bounded in SNR in multipath fading channels, even when the number of paths is finite. The lower threshold for QPSK (10 dB) versus 16-QAM (15 dB) attributes to the different susceptibility of these modulation schemes to fading - phase-based encoding of QPSK provides greater resilience against amplitude variations compared to amplitude- and phase-dependent constellation points of 16-QAM [52], [53]. This behavior suggests that higher-order modulation schemes become increasingly vulnerable to path gain variations in multipath environments, requiring higher SNR thresholds to maintain acceptable performance.

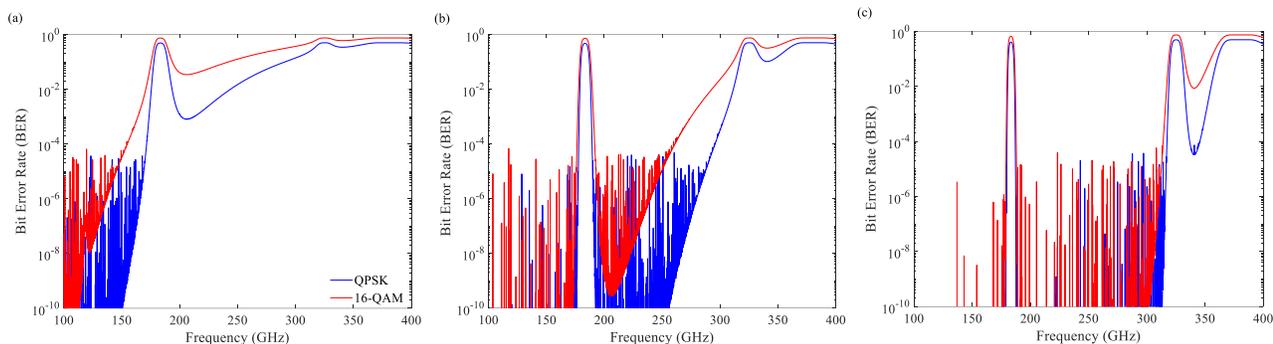

**Fig. 8.** BER performance with respect to frequency with a channel height of Hd1 under the output power of (a) 10 dBm, (b) 20 dBm and (c) 30 dBm. (b)-(c) keep the same legend with (a).

To further investigate the implications of these critical SNR thresholds, we extended our BER analysis from Eq. (6) and (9) by incorporating the free space path loss (FSPL) and atmospheric absorption [48] with Stable distribution fading model. As shown in Fig. 8, our analysis reveals a counter-intuitive relationship between transmission power and available operating frequencies. At lower transmission power levels, the system can actually operate at higher frequencies, while increasing transmission power does not necessarily improve BER performance as might be expected. This behavior creates distinct operating regimes: for lower THz frequencies, the system must either operate at reduced power levels to maintain BER above $10^{-4}$ as in Fig. 8(a), or significantly increase transmission power to achieve near error-free performance (BER = $10^{-10}$), as demonstrated in Fig. 8(c). This finding indicates that the system designers must carefully balance transmission power against desired operating frequency range, rather than simply maximizing power output.

## IV. CONCLUSION

As the deployment of terahertz communications for high-capacity wireless systems faces critical challenges from vegetation-induced channel impairments, this article aimed to develop and validate comprehensive models for characterizing these effects at frequencies from 220-230 GHz through extensive laboratory and outdoor measurements. We demonstrate that the VED model can accurately characterize vegetation-induced power loss (up to 20 dB) across different heights and seasonal conditions with remarkable accuracy while maintaining parameter stability, providing a reliable tool for link budget calculations in future THz fixed wireless access networks. This model successfully captures foliage density variations in attenuation (2.7-4.0 dB) observed between dense and sparse lilac conditions. Our statistical analysis reveals that vegetation channels at THz frequencies exhibit heavy-tailed Stable distributions rather than conventional fading models, reflecting the complex wave-matter interactions at these wavelengths.

Through experimental validation, we verify the accuracy of the BER model by measuring the EVM in laboratory conditions, establishing a quantitative bridge between channel characteristics and system-level performance metrics. Most significantly, we identify critical SNR thresholds (10 dB for QPSK, 15 dB for 16-QAM) where BER performance transitions from monotonic improvement to oscillatory behavior, a phenomenon consistent across different vegetation densities and measurement heights. This finding represents a fundamental limitation in vegetation-impaired THz channels and suggests that conventional strategies of simply increasing transmission power may not effectively improve reliability beyond certain thresholds. Instead, system designers should consider adaptive modulation and coding schemes, spatial diversity approaches and and dynamic error correction strategies that can respond to the channel's statistical nature.

Future work should focus on two critical areas identify in this article - developing specialized measurement systems to definitively identify the physical mechanisms responsible for the negative *D* value phenomenon in sparse vegetation, and conducting direct experimental validation of BER behavior using modulated data links to verify the predicted SNR thresholds and oscillatory performance under heavy-tailed fading conditions. Additional studies of vegetation types and environmental effects would further validate these findings.

12
> REPLACE THIS LINE WITH YOUR MANUSCRIPT ID NUMBER (DOUBLE-CLICK HERE TO EDIT) <[7] N. Meili, J. A. Acero, N. Peleg, G. Manoli, P. Burlando, and S. Fatichi, "Vegetation cover and plant-trait effects on outdoor thermal comfort in a tropical city," *Building and environment*, vol. 195, p. 107733, 2021.

[8] B. Li, X. Zhang, R. Wang, Y. Mei, and J. Ma, "Leaf water status monitoring by scattering effects at terahertz frequencies," *Spectrochimica acta. Part A, Molecular and biomolecular spectroscopy*, vol. 245, p. 118932, Jan 15 2021, doi: 10.1016/j.saa.2020.118932.

[9] A. Afsharinejad, A. Davy, and M. Naftaly, "Variability of Terahertz Transmission Measured in Live Plant Leaves," *IEEE Geoscience and Remote Sensing Letters*, vol. 14, no. 5, pp. 636-638, 2017, doi: 10.1109/lgrs.2017.2667225.

[10] D. Ogata, A. Sato, S. Kimura, and H. Omote, "A Study on Vegetation Loss Model with Seasonal Characteristics," *2020 14th European Conference on Antennas and Propagation (EuCAP)*, 15-20 March 2020 2020, pp. 1-4, doi: 10.23919/EuCAP48036.2020.9135532.

[11] A. N. Romanov and P. N. Ulanov, "Seasonal Differences in Dielectric Properties of Dwarf Woody Tundra Vegetation in a Microwave Range," *IEEE Transactions on Geoscience and Remote Sensing*, vol. 57, no. 6, pp. 3119-3125, 2019, doi: 10.1109/TGRS.2018.2881048.

[12] P. Horak, M. Kvicera, and P. Pechac, "Frequency Dependence of Attenuation Due to Vegetation for Satellite Services," *IEEE Antennas and Wireless Propagation Letters*, vol. 9, pp. 142-144, 2010, doi: 10.1109/LAWP.2010.2045150.

[13] A. Afsharinejad, A. Davy, P. O. Leary, and C. Brenann, "Transmission through Single and Multiple Layers of Plant Leaves at THz Frequencies," *GLOBECOM 2017 - 2017 IEEE Global Communications Conference*, 4-8 Dec. 2017 2017, pp. 1-6, doi: 10.1109/GLOCOM.2017.8254561.

[14] A. Zahid, H. T. Abbas, H. Heidari, M. Imran, A. Alomainy, and Q. H. Abbasi, "Electromagnetic Properties of Plant Leaves at Terahertz Frequencies for Health Status Monitoring," *2019 IEEE MTT-S International Microwave Biomedical Conference (IMBioC)*, May 2019, vol. 1, pp. 1-4, doi: 10.1109/IMBIOC.2019.8777782.

[15] A. Zahid et al., "Characterization and Water Content Estimation Method of Living Plant Leaves Using Terahertz Waves," *Applied Sciences*, vol. 9, no. 14, doi: 10.3390/app9142781.

[16] N. K. Periketi and A. K. Chaudhary, "Terahertz Spectral Imaging for Monitoring the Spatial and Temporal Change of Water Dynamics in Plant Leaves," *2022 Workshop on Recent Advances in Photonics (WRAP)*, March 2022, pp. 1-2, doi: 10.1109/WRAP54064.2022.9758221.

[17] B. Li, X. Zhang, R. Wang, Y. Mei, and J. Ma, "Leaf water status monitoring by scattering effects at terahertz frequencies," *Spectrochimica Acta Part A: Molecular and Biomolecular Spectroscopy*, vol. 245, p. 118932, 2021, doi: https://doi.org/10.1016/j.saa.2020.118932.

[18] A. Zahid et al., "Terahertz characterisation of living plant leaves for quality of life assessment applications," *2018 Baltic URSI Symposium (URSI)*, May 2018, pp. 117-120, doi: 10.23919/URSI.2018.8406770.

[19] Q. Jing, Z. Diao, and Z. Zhu, "Vegetation Scattering Attenuation Characteristics of Terahertz Wave," *Journal of Systems Engineering and Electronics*, vol. 34, no. 6, pp. 1501-1507, 2023, doi: 10.23919/JSEE.2022.000133.

[20] T. S. Rappaport, *Wireless Communications: Principles and Practice*, 2 ed. Cambridge: Cambridge University Press, 2024.

[21] Z. Nadir and M. Ahmad, "Pathloss Determination Using Okumura-Hata Model And Cubic Regression For Missing Data For Oman," *Lecture Notes in Engineering and Computer Science*, vol. 2181, 03/01 2010.

[22] M. A. Weissberger, "An initial critical summary of models for predicting the attenuation of radio waves by trees," 1982.

[23] Seville, A.;Craig, K.H..Semi-empirical model for millimetre-wave vegetation attenuation rates. *Electronics Letters*, 1995, Vol.31(17): 1507-1508.

[24] Adewumi, A.S.;Olabisi, O..Characterization and modeling of vegetation effects on UHF propagation through a long forested channel. *Progress In Electromagnetics Research Letters*, 2018, Vol.73: 9-16.

[25] R. Zabihi and R. G. Vaughan, "Simple Transmission Line Model from RET Results for Propagation Through Vegetation," *2018 IEEE International Symposium on Antennas and Propagation & USNC/URSI National Radio Science Meeting*, July 2018, pp. 1329-1330, doi: 10.1109/APUSNCURSINRSM.2018.8608622.

[26] A. Seville, "Vegetation attenuation: modelling and measurements at millimetric frequencies," *Tenth International Conference on Antennas and Propagation (Conf. Publ. No. 436)*, April 1997, vol. 2, pp. 5-8 vol.2, doi: 10.1049/cp:19970318.

[27] N. Savage, D. Ndzi, A. Seville, E. Vilar, and J. Austin, "Radio wave propagation through vegetation: Factors influencing signal attenuation," *Radio Science*, vol. 38, no. 5, pp. 9-1-9-14, 2003, doi: 10.1029/2002RS002758.

[28] O. Montero et al., "Attenuation of Radiofrequency Waves due to Vegetation in Colombia," *2018 IEEE-APS Topical Conference on Antennas and Propagation in Wireless Communications (APWC)*, Sept. 2018, pp. 940-943, doi: 10.1109/APWC.2018.8503671.

[29] B. D. Beelde, R. D. Beelde, E. Tanghe, D. Plets, K. Verheyen, and W. Joseph, "Vegetation Loss at D-Band Frequencies and New Vegetation-Dependent Exponential Decay Model," *IEEE Transactions on Antennas and Propagation*, vol. 70, no. 12, pp. 12092-12103, 2022, doi: 10.1109/TAP.2022.3209215.

[30] E. L. Cid, A. V. Alejos, and M. G. Sanchez, "Signaling Through Scattered Vegetation: Empirical Loss Modeling for Low Elevation Angle Satellite Paths Obstructed by Isolated Thin Trees," *IEEE Vehicular Technology Magazine*, vol. 11, no. 3, pp. 22-28, 2016, doi: 10.1109/MVT.2016.2550008.

[31] D. Li et al., "Ground-to-UAV sub-terahertz channel measurement and modeling," *Optics Express*, vol. 32, no. 18, pp. 32482-32494, 2024.

[32] J. Cui et al., "Terahertz channel modeling based on surface sensing characteristics," *Nano Communication Networks*, vol. 42, p. 100533, 2024.

[33] S. Chauhan, B. B. Manisha, K. C. Kandpal, and A. Kumar, "Analyzing preferred indoor ornamental potted plants for their air pollution tolerance ability," *Polish Journal of Environmental Studies*, vol. 31, pp. 2019-2027, 2022.

[34] P. Li et al., "Eavesdropping risk evaluation for non-line-of-sight terahertz channels by a metallic wavy surface in rain " *Journal of the Optical Society of America B*, vol. 41, no. 9, pp. 1995-2002, 2024.

[35] P. Li et al., "Scattering and Eavesdropping in Terahertz Wireless Link by Wavy Surfaces," *IEEE Transactions on Antennas and Propagation*, vol. 71, no. 4, pp. 3590-3597, 2023, doi: 10.1109/TAP.2023.3241333.

[36] L. Hsin-Piao and T. Ming-Jian, "Two-Layer multistate Markov model for modeling a 1.8 GHz narrow-band wireless propagation channel in urban Taipei city," *IEEE Transactions on Vehicular Technology*, vol. 54, no. 2, pp. 435-446, 2005, doi: 10.1109/TVT.2004.841523.

[37] W. Hong Shen and N. Moayeri, "Finite-state Markov channel-a useful model for radio communication channels," *IEEE Transactions on Vehicular Technology*, vol. 44, no. 1, pp. 163-171, 1995, doi: 10.1109/25.350282.

[38] B. D. Beelde, D. Plets, and W. Joseph, "Characterization of Vegetation Loss and Impact on Network Performance at $V$-Band Frequencies," *IEEE Antennas and Wireless Propagation Letters*, vol. 22, no. 3, pp. 596-600, 2023, doi: 10.1109/LAWP.2022.3219556.

[39] A. Subramanian, A. Sundaresan, and P. K. Varshney, "Detection of dependent heavy-tailed signals," *IEEE Transactions on Signal Processing*, vol. 63, no. 11, pp. 2790-2803, 2015.

[40] Y. S. Meng, Y. H. Lee, and B. Ng, "Study of propagation loss prediction in forest environment," *Progress in Electromagnetics Research B*, vol. 17, pp. 117-133, 01/01 2009, doi: 10.2528/PIERB09071901.

[41] Breda, Nathalie. "Ground-based measurements of leaf area index: a review of methods, instruments and current controversies." *Journal of experimental botany* 54 392 (2003): 2403-17 .

[42] G. J. U. M. Frazer and P. Documentation, "Gap Light Analyzer (GLA) Imaging software to extract canopy structure and gap light transmission indices from true-colour fisheye photographs," 1999.

[43] H. Fang, F. Baret, S. Plummer, and G. Schaepman‐Strub, "An overview of global leaf area index (LAI): Methods, products, validation, and applications," *Reviews of Geophysics*, vol. 57, no. 3, pp. 739-799, 2019.

[44] Y. Lv, X. Yin, C. Zhang, and H. Wang, "Measurement-Based Characterization of 39 GHz Millimeter-Wave Dual-Polarized Channel Under Foliage Loss Impact," *IEEE Access*, vol. 7, pp. 151558-151568, 2019, doi: 10.1109/ACCESS.2019.2945042.

[45] F. K. Schwering, E. J. Violette, and R. H. Espeland, "Millimeter-wave propagation in vegetation: experiments and theory," *IEEE Transactions on Geoscience and Remote Sensing*, vol. 26, no. 3, pp. 355-367, 1988.

[46] E. Östlin, "On Radio Wave Propagation Measurements and Modelling for Cellular Mobile Radio Networks," 2009.

[47] Z. Fang, H. Guerboukha, R. Shrestha, M. Hornbuckle, Y. Amarasinghe, and D. M. Mittleman, "Secure Communication Channels Using Atmosphere-Limited Line-of-Sight Terahertz Links," *IEEE Transactions on Terahertz Science and Technology*, vol. 12, no. 4, pp. 363-369, 2022, doi: 10.1109/tthz.2022.3178870.

[48] "International Telecommunication Union Recommendation (ITU-R) P.676-11: Attenuation by atmospheric gases." [Online]. Available: https://www.itu.int/rec/R-REC-P.676-11-201609-S/en.

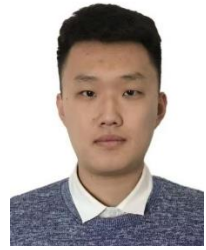
**Chenxi Wang**, born in 1998, Master, engineer. Main research interests: millimeter wave communication parameter measurement, radio measurement testing.

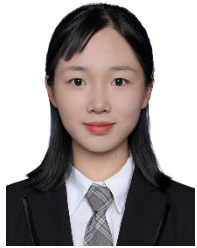
**Mingxia Zhang** received her Bachelor's degree in Electronic Information Engineering from Beijing Institute of Technology in 2024, and joined Prof. Jianjun Ma's group at Beijing Institute of Technology in 2024 to pursue a doctorate degree in Electronic Science and Technology. Her current research interests include indoor/outdoor terahertz channel measurement, modeling and terahertz detection.

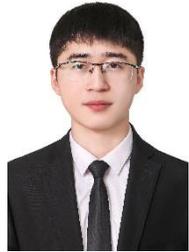
**Jiayuan Cui** graduated from Dalian Maritime University with a bachelor's degree, majoring in Electronic Information Engineering. Now he is with the Beijing Key Laboratory of Millimeter and Terahertz Wave Technology, and the School of Information and Electronics, Beijing Institute of Technology, majoring in Electronic Science and Technology, and the research direction is terahertz channel model.

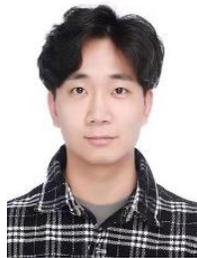
**Guohao Liu** graduated from Nanjing Normal University, majoring in electronic information. He is now studying for a master's degree in the Institute of Microwave and Terahertz, School of Integrated Circuits and Electronics, Beijing Institute of Technology, majoring in a new generation of electronic information technology, and his research field is grain moisture detection.

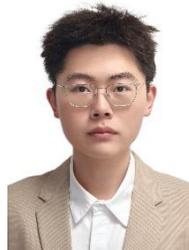
**Yuheng Song** graduated with a bachelor's degree in Integrated Circuit Design and Integrated Systems from Shandong University. He is currently pursuing a master's degree at the Institute of Microwave and Terahertz, School of Integrated Circuits and Electronics, Beijing Institute of Technology, majoring in next-generation electronic information technology. His research focuses on the measurement and modeling of indoor and outdoor terahertz channels.

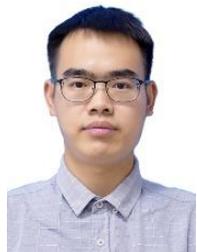
**Da Li** graduated from Shandong University with a bachelor's degree, majoring in communication engineering. Now he is with the Beijing Key Laboratory of Millimeter and Terahertz Wave Technology, and the School of Information and Electronics, Beijing Institute of Technology, majoring in New-Generation Electronic Information Technology, and the research direction is terahertz channel model.

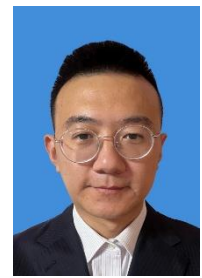
**He Jiang**, born in 1988, Doctor, senior engineer. His research interests include millimeter wave communication and channel parameter measurement, terahertz pulse and communication parameter measurement, AD hoc network communication performance test.

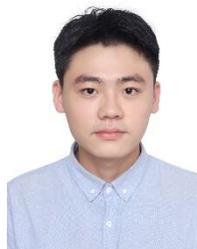
**Jiabiao Zhao** received the bachelor's degree in communication engineering from the Zhengzhou University , Henan, China, in 2021. Now he is currently pursuing the doctor's degree in the Beijing Key Laboratory of Millimeter and Terahertz Wave Technology, and the School of Integrated Circuits and Electronic, Beijing Institute of Technology, Beijing. His main research directions are terahertz indoor/outdoor propagation and link security.

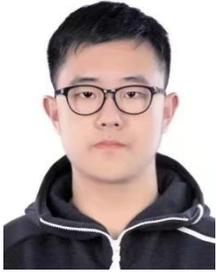

**Jiacheng Liu** received the bachelor's degree in computer science and technology from the Beijing Institute of Technology, Beijing, China, in 2022, where he is currently pursuing the master's degree in electronic science and technology. His current research focus primarily lies at the intersection of terahertz indoor and outdoor transmission and deep learning.

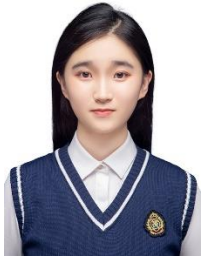

**Yue Su** graduated from Hebei Normal University, majoring in information security. She is now studying for a master's degree in the Institute of Microwave and Terahertz, School of Integrated Circuits and Electronics, Beijing Institute of Technology, majoring in a new generation of electronic information technology, and her research field is diffraction of terahertz signals.

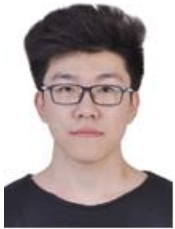

**Wenbo Liu** graduated from University of Central Lancashire with a bachelor's degree and Coventry University with a master's degree. Now he is with the Beijing Key Laboratory of Millimeter and Terahertz Wave Technology, and the School of Integrated Circuits and Electronic, Beijing Institute of Technology.

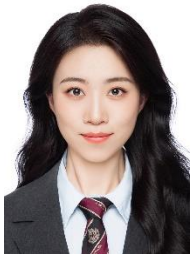

**Peian Li** holds a bachelor's degree from Minzu University of China, a master's degree from Beijing Jiaotong University, and a doctor of engineering degree from Beijing Institute of Technology. She is currently a postdoctoral research fellow at the School of Integrated Circuit and Electronics, Beijing Institute of Technology. Her research focuses on terahertz channel characteristics and physical layer security, terahertz wireless communication, interactions between terahertz waves and plasma.

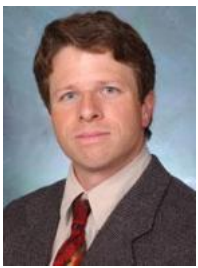

**Daniel M. Mittleman** (Fellow, IEEE) received his B.S. in physics from the Massachusetts Institute of Technology in 1988, and his M.S. in 1990 and Ph.D. in 1994, both in physics from the University of California, Berkeley, under the direction of Dr. Charles Shank. He then joined AT&T Bell Laboratories as a post-doctoral member of the technical staff, working first for Dr. Richard Freeman on a terawatt laser system, and then for Dr. Martin Nuss on terahertz spectroscopy and imaging. Dr. Mittleman joined the ECE Department at Rice University in September 1996. In 2015, he moved to the School of Engineering at Brown University. His research interests involve the science and technology of terahertz radiation. He is a Fellow of the OSA, the APS, and the IEEE, and is a 2018 recipient of the Humboldt Research Award. In 2018-2020, he served a three-year term as Chair of the International Society for Infrared Millimeter and Terahertz Waves, and received the Society's Exceptional Service Award in 2022. In 2023-2025, he is a Mercator Fellow of the Deutsche Forschungsgemeinschaft (DFG), in affiliation with the Meteracom project.

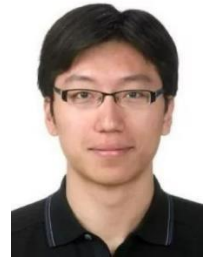

**Fei Song** (Member, IEEE) is a Full Professor with the National Engineering Research Center for Advanced Network Technology and the School of Electronic and Information Engineering, Bei jing Jiaotong University, Beijing, China. His current research interests include network architecture, system security, protocol optimization, and cloud computing.

Prof. Song serves as a Technical Reviewer for several journals including, IEEE Communications Magazine, IEEE Internet of Things Journal, IEEE Transactions on Industrial Informatics, IEEE Transactions on Services Computing, and IEEE Transactions on Emerging Topics in Computing.

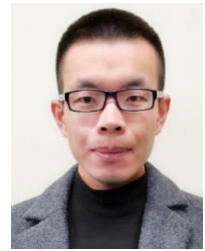

**Jianjun Ma** (Member, IEEE) was born in Qingdao, China, in 1986. He received the Ph.D. degree in applied physics from New Jersey Institute of Technology, Newark, NJ, USA and in 2015, under the guidance of Prof. John F. Federici. His Ph.D. dissertation was about the weather impacts on outdoor terahertz and infrared wireless communication links.

In 2016, he joined Prof. Daniel M. Mittleman's group at Brown University, RI, USA, as a postdoctoral research associate. In 2019, he joined Beijing Institute of Technology, Beijing, China, as a professor. His current research interests include terahertz and infrared wireless communications, terahertz time domain detection, terahertz waveguides, terahertz indoor/outdoor propagation and link security.